# CONTROL SYSTEM FOR ELECTROMAGNET POWER SUPPLIES

E.Y.Ermolov, V.R.Kozak, E.A.Kuper, A.S.Medvedko, S.P.Petrov, V.F.Veremeenko, BINP, Novosibirsk, Russia

Abstract

A set of power supplies (PS) with output power rated from 100 W up to 10 KW for electromagnets was developed. These PS have range of current tuning of 60-80 db with high accuracy (error should be less than 0,01%). Some types of power supplies have bipolar output current. The report will describe a set of unified embedded devices for control and measurements of PS incorporated into distributed control systems. These embedded devices includes DAC, ADC with multiplexers and status input/output registers. The devices include microprocessor and allow work with minimal host involving. This capability decreases traffic, provides synchronous tuning of all electromagnet elements of particle accelerator and increases reliability of entire system.

## 1 INTRODUCTION

Budker Institute of Nuclear Physics (BINP) is building a few installation (VEPP-5, FEL) and upgrading a existing collider (VEPP-2000, VEPP-4 [1]). Mentioned installations include hundreds of magnet elements (lenses, correctors) which are powered by controlled power supplies designed and produced by BINP. All power supplies are computer controlled via digital-to-analog converters. Current and voltage in elements are measured with analog-to-digital converters.

There was developed a set of embedded devices for control of power supplies. They have unified implementation, similar connectors and software protocol. Integration with control system is provided by popular network- CANBUS.

## 2 TECHNICAL REQUIREMENTS

Magnet elements and power supplies, correspondingly, may be divided into two different groups. A first group includes bending magnets and lenses. A second one includes low current correctors. Elements of first group require high stability and accuracy of power supplies. Typical requirements to accuracy of this power supplies are from 0,01% to 0,001%. For corrector's power supplies these requirements are lower. It ranges approximately near 0,1% typically.

A wide gap in required accuracy makes us to design a two different type of controllers. The first is precise controller for bending magnet and lens power supplies. The second is inexpensive multi-channel device for multi-channel power supply systems for correctors.

A precise controller should provide one DAC channel with resolution approximately 0,001% and accuracy at least 0,01%. It should provide also 4-5 channels of ADC with similar accuracy and resolution of measurement of the output current and voltage. It should also measure of some additional voltages inside PS. It is very useful to check logical status of some points of power supplies (protection circuitry status, presence of water flow and etc) and ON/OFF control for different purposes. For this tasks the controller should provide a few channels of input/output register with optocouplers.

Inexpensive multi-channel controller should provide multiple-channel DAC (we have chosen 16 channels per controller) with accuracy at least 0,1 % and multi-channel ADC (we have chosen 40 channels per controller) with the same accuracy. For digital IN/OUT it should include a few channels of input/output register with optocouplers.

For connection of the controllers with control computers we have chosen CANBUS. It provides:
- high reliability;
- determined delivery time for high priority packets;
- wide support by chip and board manufacturers;
- opportunity of galvanic isolation between controller and media;
- growing popularity of this network in world physical centers.

There are a number of CAN application level protocols. A structure and resources of chosen microprocessor don't prevent to implement one of them in our devices. But using of many CAN-devices requires to reduce traffic as much as possible. This reason lead us to direct use of the CAN message objects. But this opportunity may be realized later.

## 3 STRUCTURE OF CONTROLLERS

A multi-channel controller is implemented in two physically independent devices. One of them contains multi-channel DAC, the second includes multi-channel ADC. Other properties of these devices are identical.

Below in Fig.1 is presented a block-diagram of all mentioned devices. All of them includes a few identical parts (microprocessor, CAN controller, DC-DC converter, input/output register) and specific analog parts, which are different for both devices.

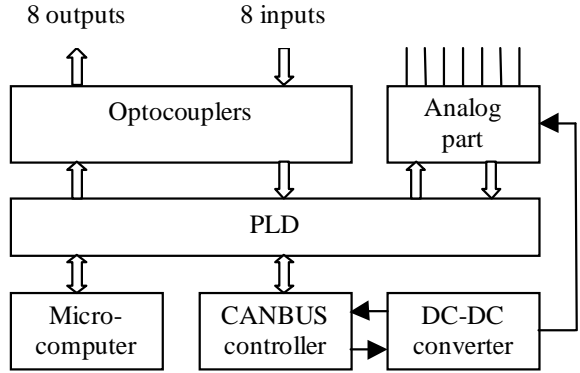

Figure 1: Block-diagram of embedded controller.

## 4 DESING OF CONTROLLERS

Analog part of precise controller (CDAC20) incorporates one channel of DAC and multi-channel ADC. A structure of analog part is shown in Fig.2.

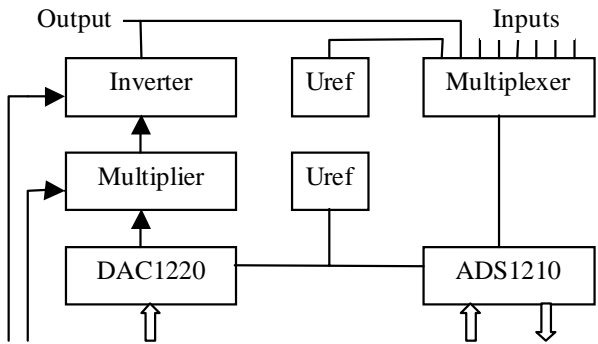

Fig.2. Analog part of precise controller.

A design is based on Burr-Brown delta-sigma converters. A design of ADC is quite classical. Input signals connected with ADS1210 through multiplexer with incorporated overvoltage protection (Burr-Brown MPC507). 5 two-wired channels intended for external signal measurements. One channel is connected with DAC output inside board. This feature gives us opportunity to control DAC and to correct it by microprocessor. One channel of multiplexer is connected with "ground" and another – with reference voltage. This feature allows using "system calibration" of ADC chip and to reduce most sources of inaccuracy.

Design of DAC have some specific. Chosen chip (DAC1220) provides output signal from 0 to +5V. Most our power supplies require range up to 10 V. So, DAC1220 voltage is doubled by circuitry with "flying capacity", then it may be inverted by similar circuitry.

The analog part of precise controller is galvanically isolated from digital part of device. It is provided with DC-DC converter and optocouplers. Galvanic isolation allows us to improve quality of signal transfer between controller and power supply.

An analog part of multi-channel DAC (CANDAC16) is based on single DAC chip (AD669, Analog Devices) and 16 sample-and-hold devices (see Fig.3).

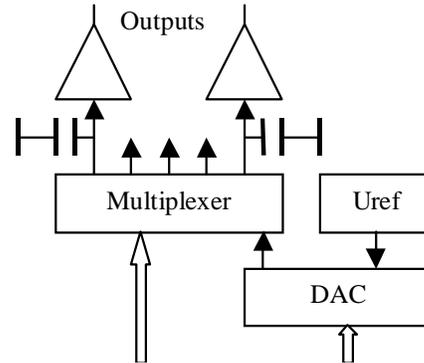

Fig.3. Analog part of multi-channel DAC.

Microprocessor loads code into DAC chip, then connect output of DAC with chosen capacitor to update charge. After some time microprocessor repeats this sequence for next channel. This approach allows built a multi-channel DAC with low cost per channel. We use external reference source to improve stability and accuracy of device. There is no here of the galvanic isolation between analog and digital areas. It makes design more cheap.

An analog part of multi-channel ADC (CANADC40) is based on sigma-delta ADC chip (ADS1210, Burr-Brown) (see fig.4).

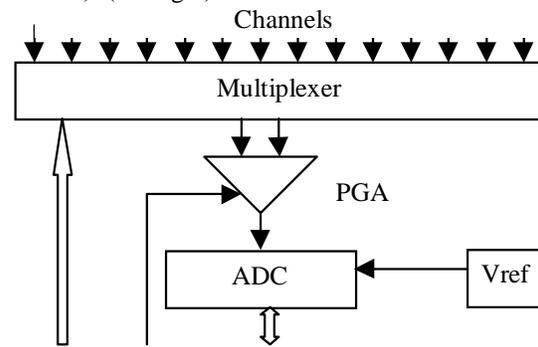

Fig.4 Analog part of multi-channel ADC.

This circuitry contains 40-channel multiplexer based on MPC507, (multiplexer with incorporated over-voltage protection), programmable gain amplifier (PGA), delta-sigma ADC (ADS1210, Burr-Brown) and reference source. A PGA chip is used to allow automatic changing of the input range of meter. Here isn't galvanic isolation between ADC analog and digital areas.

## 5 EMBEDDED SOFTWARE

Embedded software for all devices is written in assembler to ensure determined time for critical sections of code. All devices use very similar or identical messages.

Controllers with DAC on board have additional feature. They can load from line a few files, which describe changing output voltage in time. Host computer can request to process these files by single, group or all devices in network. This feature allows to reconfigure all magnet system (or part of it) of accelerator synchronously. A file contains a set of points and embedded software calculates intermediate points during processing this file. This calculation use a method of linear interpolation.

## 6 PARAMETERS

Below is presented parameters of designed devices.

CDAC20- precise controller.

| DAC accuracy | 0,005% |
|---|---|
| DAC resolution | 21 bits |
| DAC channels number | 1 |
| ADC accuracy | 0,003% |
| ADC resolution | 24 bits |
| ADC channels number | 5+3 |
| ADC and DAC range | ±10V |
| Channels of output register | 8 |
| Channels of input register | 8 |

CANDAC16- multi-channel DAC.

| DAC accuracy | 0,05% |
|---|---|
| DAC resolution | 16 bits |
| DAC channels number | 16 |
| DAC range | ±10V |
| Channels of output register | 8 |
| Channels o input register | 8 |

CANADC40- multi-channel ADC.

| ADC accuracy | 0,03% |
|---|---|
| ADC resolution | 24 bits |
| ADC channels number | 40 |
| ADC range | ±10V |
| Channels of output register | 8 |
| Channels of input register | 8 |

## CONCLUSION

A set of embedded controllers for power supplies was designed and tested. A number of controllers is used in power supply system of VEPP-5, FEL. Using of the embedded controllers has minimized interconnection, simplified maintenance and reduced commissioning time. We suppose to extend this approach to other large installation of institute.